\documentclass[sigconf]{acmart}
\usepackage{stfloats}
\AtBeginDocument{%
  }

\setcopyright{acmlicensed}
\copyrightyear{2026}
\acmYear{2026}
\acmDOI{XXXXXXX.XXXXXXX}
\acmConference[GLSVLSI '26]{Great Lakes Symposium on VLSI}{June 24--26,
  2026}{Finger Lakes, NY}
\acmISBN{978-1-4503-XXXX-X/2026/06}

\begin{document}

% \title{The Name of the Title Is Hope}
\title{DarkFlow: Hierarchical Digital SiPM Architecture with Low-Loss Dataflow Readout for Dark Matter Detection}


% \author{Author names}
% \affiliation{%
%   \institution{Author affiliation}
%   \city{City}
%   \country{Country}
%   }
% \email{Contact info}

\author{Zirui Wang}
\orcid{0009-0009-2368-6272}
\email{zwang829@ucr.edu}
\affiliation{%
  \institution{University of California, Riverside}
  \city{Riverside}
  \country{USA}}

\author{Aras Repond}
\orcid{0000-0003-1338-8780}
\email{aras.repond@email.ucr.edu}
\affiliation{%
  \institution{University of California, Riverside}
  \city{Riverside}
  \country{USA}}

\author{Shawn Westerdale}
\orcid{0000-0001-8824-6205}
\email{shawn.westerdale@ucr.edu}
\affiliation{%
  \institution{University of California, Riverside}
  \city{Riverside}
  \country{USA}}

\author{Wantong Li}
\orcid{0000-0002-8288-393X}
\email{wantong.li@ucr.edu}
\affiliation{%
  \institution{University of California, Riverside}
  \city{Riverside}
  \country{USA}}

\renewcommand{\shortauthors}{Wang et al.}

\begin{abstract}
Direct dark matter detection experiments require large-scale photon sensing arrays with hundreds of thousands of synchronized readout channels. Silicon photomultipliers (SiPMs) have emerged as a leading candidate for these detectors due to their high integration density, low bias voltage, and superior radiopurity. However, existing digital SiPM readout architectures struggle to simultaneously preserve nanosecond-level temporal resolution for sparse scintillation events and sustain data integrity during high-intensity photon bursts. We propose DarkFlow, a hierarchical digital SiPM architecture that features local data aggregation, compact relative-time encoding, consumer-driven backpressure, and occupancy-aware eDRAM burst buffering within a unified dataflow framework. We show that DarkFlow maintains ultra-low packet loss at billion-photon event rates, where conventional architectures can exceed 80\% data loss. Besides, the occupancy-aware refresh achieves a 2.14$\times$ improvement in effective refresh rate over conventional global refresh. Hardware evaluation in GlobalFoundries 22nm node confirms that the digital readout datapath accounts for less than 0.86\% of the detector area and complies with the strict power budget in liquid argon environments.

\end{abstract}





\maketitle

\section{Introduction}

Resolving the nature of dark matter remains a central challenge in particle physics. 
Current direct detection experiments search for weakly interacting massive particles (WIMPs) by capturing the faint scintillation photons generated during rare WIMP-nucleus collisions~\cite{canci2025darkside}.
Dual-phase liquid argon time projection chambers (LAr-TPCs), which measure prompt scintillation pulses followed by delayed electroluminescence pulses induced by ionization electrons, are widely used in particle physics experiments. To guarantee efficient photon detection across the full detector, dark matter detectors require an extensive and uniformly controlled optical sensing surface. This presents a high-density parallel single-photon counting problem that demands low-power, high-bandwidth readout architecture with ultra-low data loss.

Traditional photomultiplier tubes used as dark matter detectors are fundamentally constrained by their bulky form factors, kV-level bias requirements, and intrinsic radioactivity~\cite{acciarri2012demonstration}.
To achieve high radiopurity and superior photon detection efficiency, silicon photomultipliers (SiPMs) composed of dense arrays of single-photon avalanche diodes (SPADs) have been proposed as a promising alternative.
SiPMs offer superior radiopurity, high photon detection efficiency, and the high-density integration potential essential for scaling to hundreds of thousands of channels~\cite{rogers2024production}.

\begin{figure}[!b]
  \centering
  % \vspace{-15pt}
  \includegraphics[width=\linewidth]{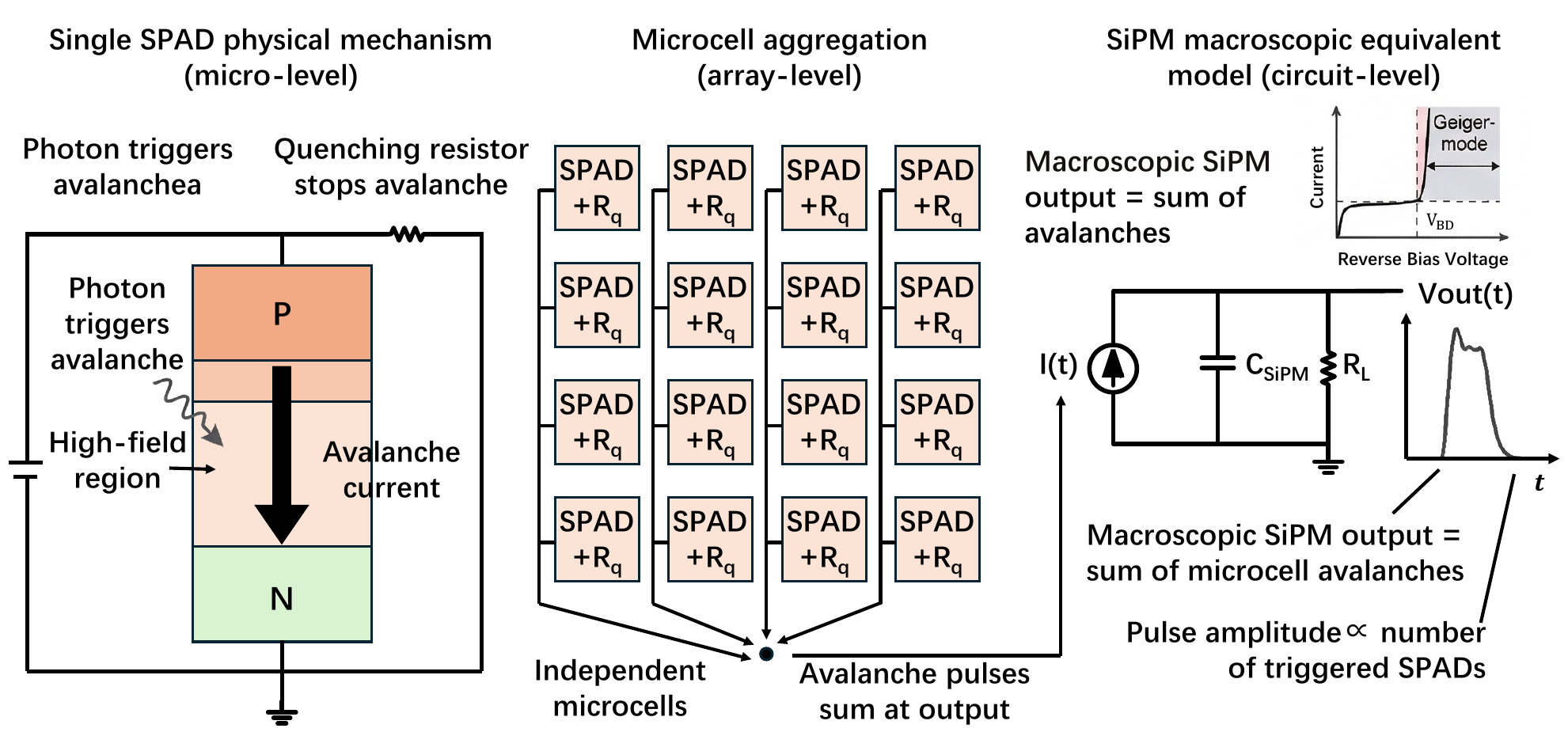}
  \caption{Physical mechanism of a SPAD and its macroscopic aggregation in a SiPM.}
  \label{fig:introduction}
  \Description{This is a diagram showing introduction of the Digital SiPM System.}
\end{figure}

\begin{figure*}[!htbp]
  \centering
 \includegraphics[width=0.90\textwidth]{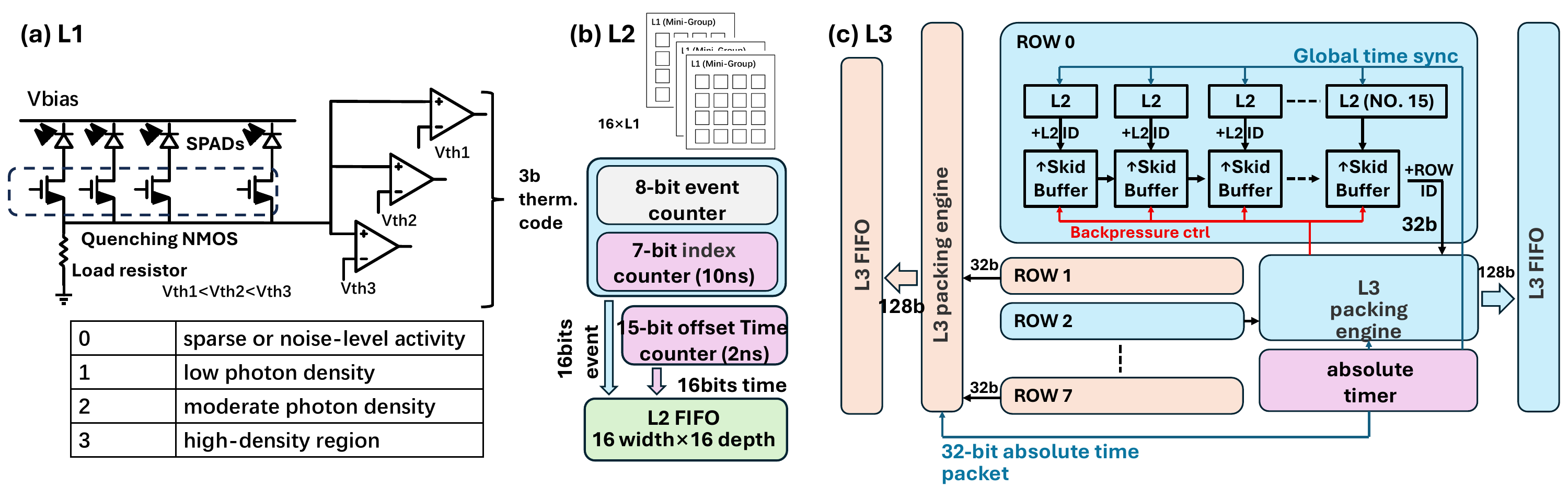}
  \caption{Top-level DarkFlow architecture, showing the hierarchy of (a) L1 sensing unit, (b) L2 pixel tile, and (c) L3 row-level aggregation and packet collection.}
  \vspace{-10pt}
  \label{fig:architecture}
\end{figure*}

However, the large-scale integration of SiPMs within a TPC environment introduces conflicting design requirements. On one hand, the design must specifically resolve the conflict between two signal modalities: S1 and S2. The prompt S1 scintillation (ns-level fast component) requires fine timing resolution for pulse shape discrimination to reject background noises~\cite{deap2020liquid}. Conversely, the S2 electroluminescence ($\mu$s-level slow component) demands fine-grained mapping to identify multi-scatter topologies~\cite{agnes2018electroluminescence}. 
Due to its high-gain nature, S2 generates a peak throughput exceeding 10 Gbps. On the other hand, the system must operate under extreme physical constraints: frontend electronics are co-located with the SiPM sensors and impose a strict mW-level power budget per channel to prevent liquid argon boiling~\cite{consiglio2020cryogenic}. Furthermore, scaling up the uplink bandwidth is physically prohibitive. Transmitting high-bandwidth data across long cables requires power-hungry I/O interfaces that violate the local thermal budget, while the dense cabling itself conducts unwanted ambient heat directly into the liquid argon~\cite{10540654}.

As a result, the readout architecture for large-scale SiPM arrays must simultaneously resolve the S1/S2 bandwidth conflict, scale to high channel counts within a tightly constrained power and I/O budget, and maintain robust data integrity under extreme photon burst conditions. This work addresses these challenges through the DarkFlow architecture. Our main contributions are as follows:

\noindent\textbullet ~~~ We propose DarkFlow, a hierarchical SiPM dataflow readout architecture co-designed for sparse S1 and burst-heavy S2 traffic patterns in large-scale dark matter detectors, maintaining ultra-low data loss at billion-photon event rates.

\noindent\textbullet ~~~ We design a compact hierarchical packet format combined with burst buffering and occupancy-aware refresh that minimizes per-event bandwidth and maximizes on-detector data retention, scaling to high channel counts within power and bandwidth constraints.

\noindent\textbullet ~~~ We develop an evaluation framework that quantifies digital packet loss and photon-equivalent information preservation, and we demonstrate DarkFlow's congestion robustness and lightweight overhead compared to existing readout schemes.

\section{Background}

The architectural trade-offs in particle detectors are fundamentally dictated by the underlying photon-sensing mechanism. As illustrated in Figure~\ref{fig:introduction}, the basic sensing unit in SiPM-based dark matter detector is a SPAD biased above its breakdown voltage. Upon absorbing an incident photon, the SPAD triggers a localized macroscopic current avalanche~\cite{otte2006silicon}. A SiPM scales this mechanism by connecting thousands of such microcells in parallel, transforming discrete single-photon hits into a collective macroscopic current pulse.

How these avalanche events are extracted and processed defines the readout architecture.
Traditional analog SiPMs aggregate avalanche currents macroscopically~\cite{zhao2022analog}. Large-scale integration of analog SiPMs introduces prohibitive terminal parasitic capacitance, which degrades the signal-to-noise ratio, compromises signal rise times, and sacrifices fine spatial resolution.
%\textbf{TDC-based digital SiPMs.} 

To bypass analog integration bottlenecks, digital SiPMs utilize in-pixel digitization via high-resolution time-to-digital converters~\cite{frach2010digital}. Although this approach is effective for sparse photon detections, it exhibits a fundamental architectural mismatch with high-flux signal environments. Specifically, the per-event generation and routing of multi-bit high-resolution timestamps for every SPAD trigger leads to excessive data rates that can overwhelm the local buffering capacity.
%\textbf{Asynchronous event-driven schemes.} 
Address event representation (AER) protocols offer power-efficient readouts for sparse events~\cite{9109635}, but they suffer from arbitration collisions under continuous high-flux photon bursts. Without robust backpressure mechanisms, their frontend buffers can rapidly overflow and experience severe dead-times and information loss~\cite{fischer2025design}.
These limitations collectively motivate a new readout architecture that can address the fundamental gap between large sensor arrays and limited bandwidth. 

\section{Proposed DarkFlow SiPM Architecture}

\subsection{Hierarchical Readout Architecture}

In massive sensor arrays, a global readout scheme is highly susceptible to bandwidth saturation and event pile-up during intense photon bursts. Segmented hierarchical readout mitigates this issue by localizing the initial data aggregation and temporal stamping, thus preventing the global bus from instantaneous data overflow while significantly reducing the dynamic power on long interconnects~\cite{vinayaka2019segmented}. Following this design philosophy, we design the DarkFlow readout topology as a three-level hierarchy, as shown in Figure~\ref{fig:architecture}. 
The first level (L1) serves as the foundational sensing unit, comprising a 4×4 SPAD array. As shown in Figure~\ref{fig:architecture}(a), the L1 module sums the output signals of the 16 SPADs in the analog domain and quantizes this local sum using three comparators with distinct monotonic voltage thresholds ($V_{th1} < V_{th2} < V_{th3}$). This generates a 3-bit thermometer code, which is then mapped to a compact 2-bit digital energy level. 
This front-end encoding is a transport trade-off that preserves burst structure and relative local intensity trends under constrained bandwidth.
The lowest comparator threshold ($V_{th1}$) establishes a programmable photon threshold ($P_{th}$). For standard operation, setting $P_{th}=1$ ensures absolute sensitivity to sparse single-photoelectron (1 PE) physical events, such as faint S1 scintillations. Because $P_{th}$ can be adjusted locally, setting $P_{th}=2$ on select L1 nodes acts as a targeted spatial filter. 
This suppresses defective SPADs with high dark count rates while preserving the node's ability to detect multi-photon events.
Furthermore, this programmability offers a critical system-level flow control mechanism. During massive S2 avalanches where the external I/O bandwidth is severely bottlenecked, $P_{th}$ can be globally elevated. This dynamic thresholding drops low-intensity peripheral hits to prioritize brighter signals, reducing buffer overflow risk while preserving the spatial distribution of the strongest signals. Meanwhile, the higher thresholds ($V_{th2}$ and $V_{th3}$) classify these analog sums into low, moderate, or high local photon densities to preserve the energy granularity.

\begin{figure}[t!]
  \centering
  \includegraphics[width=0.90\linewidth]{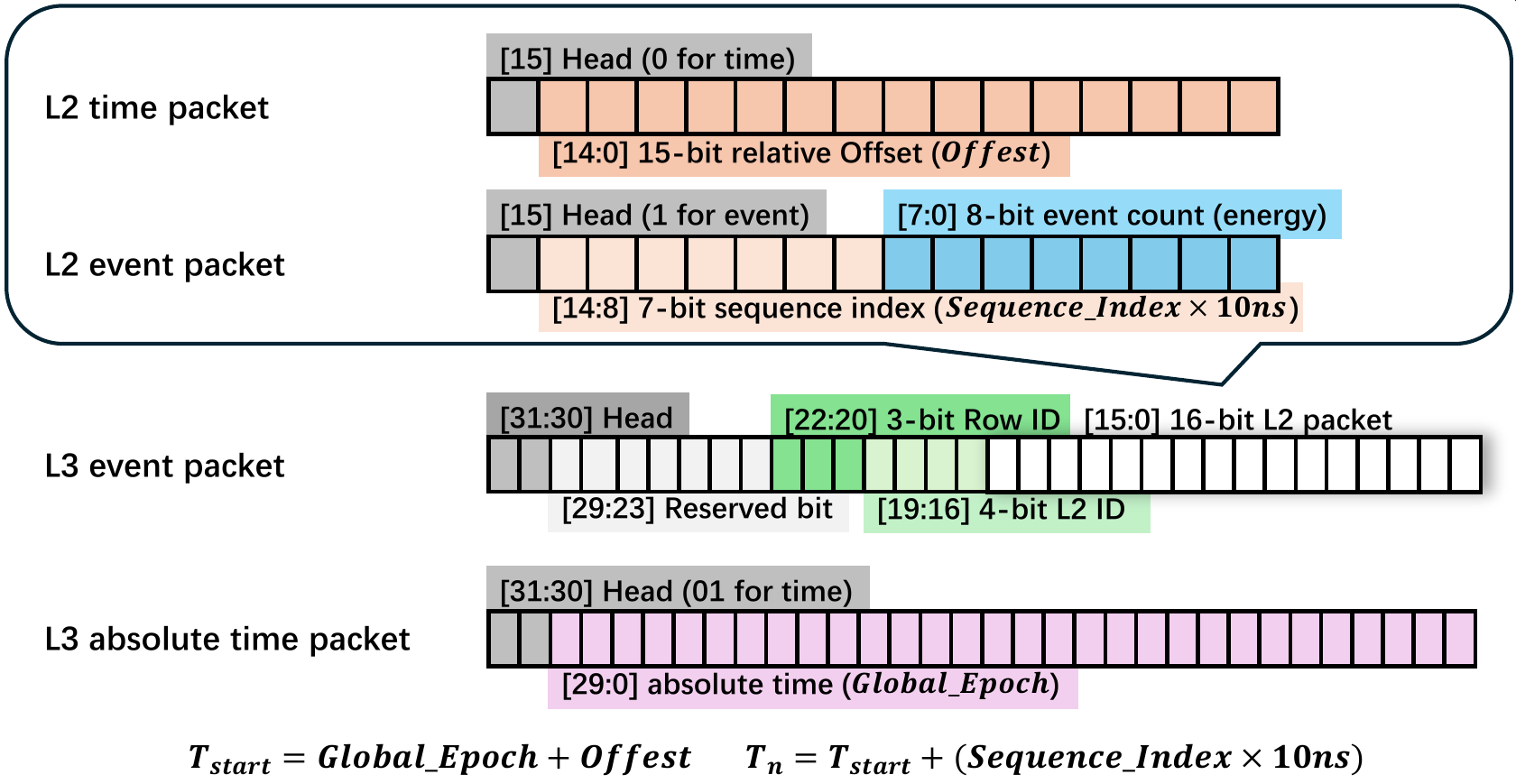}
  \caption{DarkFlow's hierarchical packet definitions.}
  \vspace{-10pt}
  \label{fig:datapacket}
\end{figure}

\begin{figure*}[b!]
  \centering
 \includegraphics[width=0.99\textwidth]
  {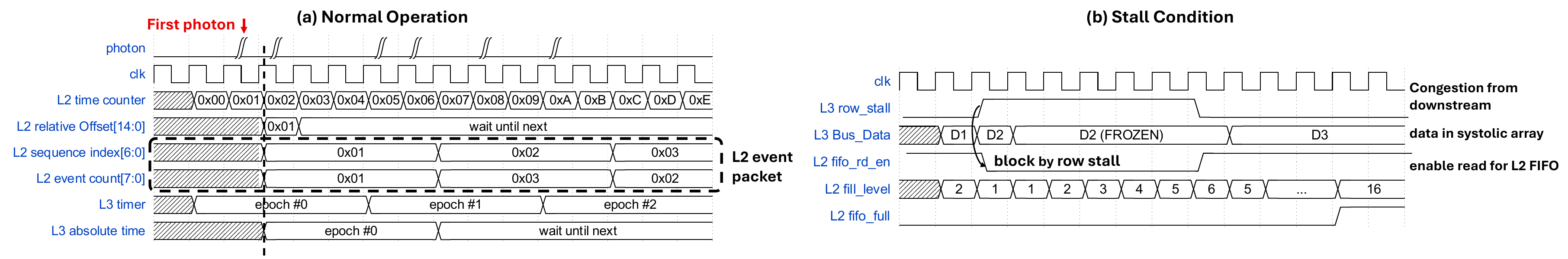}
  \caption{Timing diagram illustrating (a) the first-photon arrival, and (b) readout response under stall condition.}
  %\vspace{-15pt}
  \label{fig:waveform}
\end{figure*}

To localize data aggregation and temporal processing, the second level (L2) integrates 16 such L1 modules, as shown in Figure~\ref{fig:architecture}(b). Alongside collecting the quantized energy states from its L1 blocks, each L2 node captures the arrival time of the first photon hit with a 2-ns resolution. This high temporal precision is essential for accurately resolving the prompt leading edge of the S1 scintillation pulse. Once triggered, the subsequent spatial-energy aggregation operates continuously at a coarser 10-ns frame rate.
To overcome the routing delays and timing challenges inherent in large-scale arrays, the L2 nodes within the same row employ a systolic readout mechanism. Shown in Figure~\ref{fig:architecture}(c), nodes transmit locally buffered data sequentially to the end of the row in a daisy-chain fashion, where each transmission stage incorporates a single-slot skid buffer. Compared to a centralized arbitration scheme~\cite{9109635}, this approach eliminates long interconnects and incorporates distributed buffering to avoid data loss during data bursts. 
The third level (L3) introduces a dual-bank systolic array aggregation engine. Compared to conventional approaches that rely on shared buses or star topologies which suffer from congestion~\cite{11284747}, our L3 architecture divides the L2 systolic rows into even and odd groups. It is managed independently by left and right banks for parallel data aggregation and packing. This separation distributes the data flow across two independent paths, which allow the system to process multiple packets concurrently within one clock cycle.

\vspace{-2pt}
\subsection{Data Packet Design and Aggregation}

Conventional synchronization schemes append an absolute timestamp to every photon event. During sustained S2 bursts, this per-event timestamping introduces unacceptable bandwidth overhead and triggers packet congestion. To address this, DarkFlow separates absolute timing anchors from high-frequency local event data through the hierarchical packet format shown in Figure~\ref{fig:datapacket}.
The physical data packets generated at all L2 nodes use a compact 16-bit format, as illustrated in the timing diagram of Figure~\ref{fig:waveform}(a). Upon initial photon detection, the packet carries a 15-bit relative time offset that records the arrival time at 2-ns resolution. For subsequent events, a 7-bit sequence index tracks temporal steps at 10-ns intervals, while an 8-bit event counter records the photon count within each 10-ns window. To anchor these relative timestamps to an absolute reference, the L3 module periodically broadcasts a 32-bit global time packet that establishes a 10-$\mu$s synchronization window. To preserve spatial information, the L2 skid buffer and L3 packing engine append topological identifiers (Row ID and L2 Node ID, corresponding to the $Y$ and $X$ coordinates on the physical sensor plane) to the headers of aggregated data words before writing them to the edge FIFOs. This format reduces per-event timestamp overhead by over an order of magnitude compared to per-photon absolute timestamping, while retaining both the temporal and spatial origin of every recorded event.

During aggregation, packets from different L2 nodes compete and interleave, creating variable queuing delays. However, because all timestamps are generated at the L2 source at the moment of photon detection, the recorded timing is not affected by downstream transport delays. During offline event reconstruction, the host-side data acquisition system groups incoming packets by their spatial identifiers and reconstructs the absolute arrival time by combining each packet's relative offset and sequence index with the latest global time anchor. A sort by timestamp then restores the original event sequence regardless of packet reordering during transport.

Without flow control, the bandwidth mismatch between internal S2 data rates and the external link can lead to buffer saturation and destroy the spatial information needed for event discrimination. DarkFlow addresses this through a consumer-driven backpressure mechanism; its functional timing is captured in Figure~\ref{fig:waveform}(b), showing how stall signals propagate upstream to pause the L2 dataflow until the buffer backpressure is entirely relieved. When downstream consumers pause reception, the L3 packing engines stop writing to the edge FIFOs. Once their internal buffers are filled, the L3 module asserts a row stall signal that propagates backward to the corresponding L2 row, freezing the systolic bus and preventing L2 nodes from draining their local FIFOs. The single-slot skid buffers hold any in-flight packets on the stalled bus. Meanwhile, the 16-depth local FIFOs within L2 nodes continue to absorb new events from the analog frontend. The L1 state machine halts packet generation only when the L2 FIFO is fully saturated. This layered backpressure mechanism allows congestion to propagate from L3 back to L1 without data loss, thus sustaining high throughput during bursts without dropping packets.

\begin{figure}[t!]
  \centering
  \includegraphics[width=0.9\linewidth]{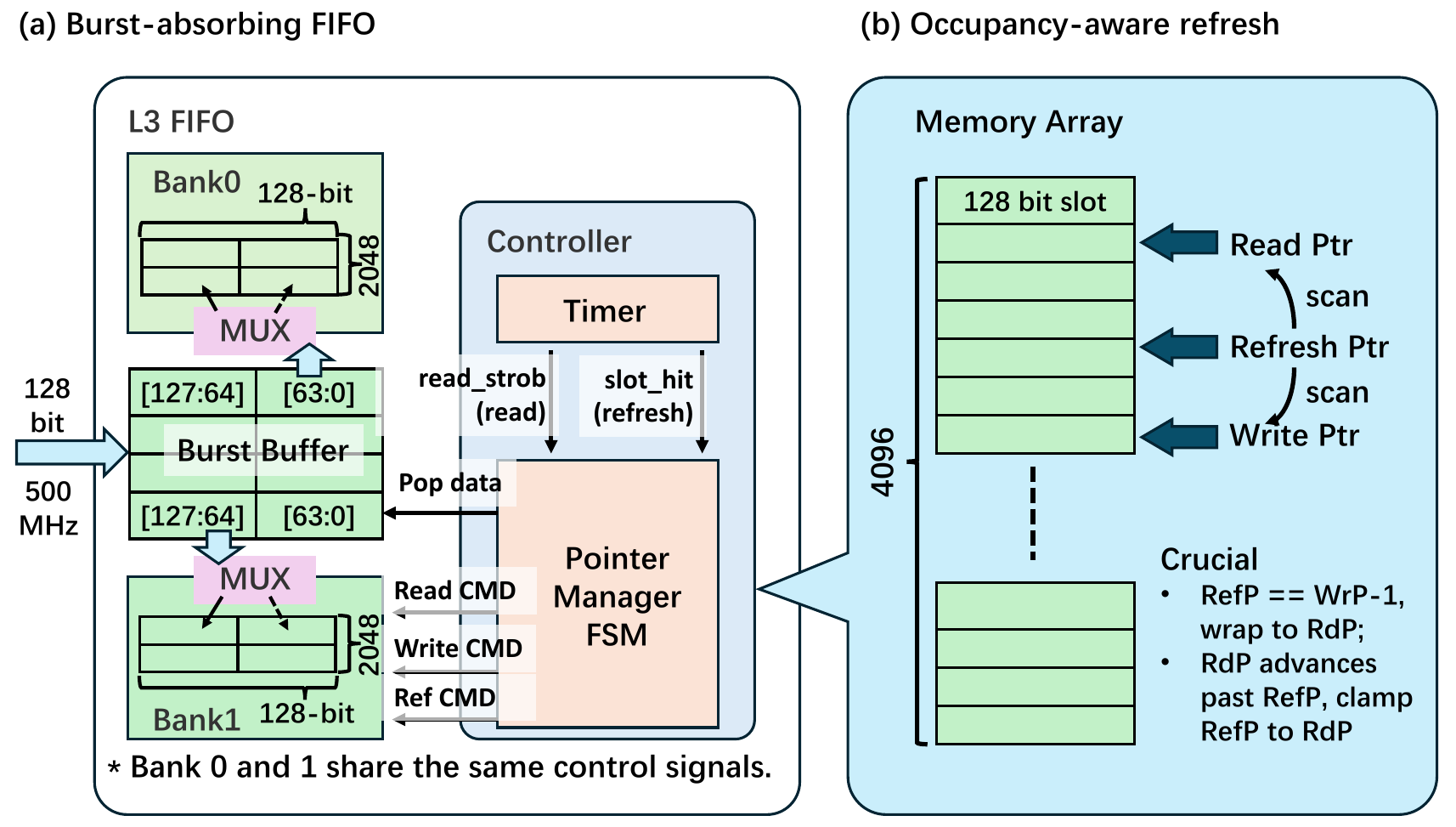}
  \caption{(a) Structure of the proposed burst-absorbing FIFO. (b) Occupancy-aware refresh mechanism.}
  \vspace{-6pt}
  \label{fig:FIFO}
\end{figure}

\subsection{Burst-Absorbing FIFO with Occupancy-Aware Refresh}

Beyond localized flow-control mechanism, addressing the system-level bottleneck also requires massive on-detector storage. Specifically, the system faces a severe bandwidth discrepancy between the restricted external interface and internal S2 data rates. To resolve this, we employ a 128-kB burst-absorbing FIFO at the L3 boundary shown in Figure~\ref{fig:FIFO}(a). 
The 128-kB capacity corresponds to a 256-bit $\times$ 4096-depth FIFO and is sized to absorb the peak packet backlog accumulated during the modeled worst-case S2 burst before the external link drains the queue.
The FIFO uses embedded DRAM (eDRAM) technology for high-density buffer. Instead of a traditional centralized memory macro, the FIFO employs two-level physical partitioning to maximize throughput and fit the narrow peripheral footprint. The FIFO architecture is bifurcated to the left and right peripheries of the sensor array (each configured as $128\text{-bit} \times 4096$), alternating to store even and odd data rows. Each side is further partitioned into top and bottom banks (each 128 columns $\times$ 2048 rows) to satisfy tall and thin floorplan constraints of SiPM designs and halve the bitline parasitic capacitance. Finally, each incoming 128-bit packet is bifurcated, with the resulting segments concurrently routed to the top and bottom banks. This enables the two partitions to share a unified control logic.
% These internal banks employ 2:1 column multiplexer to match the sensor pitch, concurrently producing 64 bits per bank that are concatenated to form the local 128-bit bus. 

During high-flux bursts, the L3 memory controller must dedicate nearly all bus bandwidth to writing incoming data and reading out to the external link. To maximize data throughput, the system dynamically suspends background refresh operations during massive S2 events. However, constrained by short eDRAM data retention times, prolonged suspension introduces a risk of data corruption. A conventional timer-based global refresh scans the entire address space sequentially, including addresses that hold no valid data. When refresh opportunities are limited during sustained bursts, this approach wastes scarce refresh cycles on empty locations. To improve refresh efficiency, we propose an occupancy-aware refresh mechanism detailed in Figure~\ref{fig:FIFO}(b). This strategy confines the custom refresh pointer within the active data window bounded by the read and write pointers, ensuring that every refresh cycle targets a valid address and achieves a 100\% data hit rate. Therefore, the eDRAM controller uses every available refresh cycle on valid data.

\section{Evaluation and Results}

\begin{figure}[t!]
  \centering
  \includegraphics[width=\linewidth]{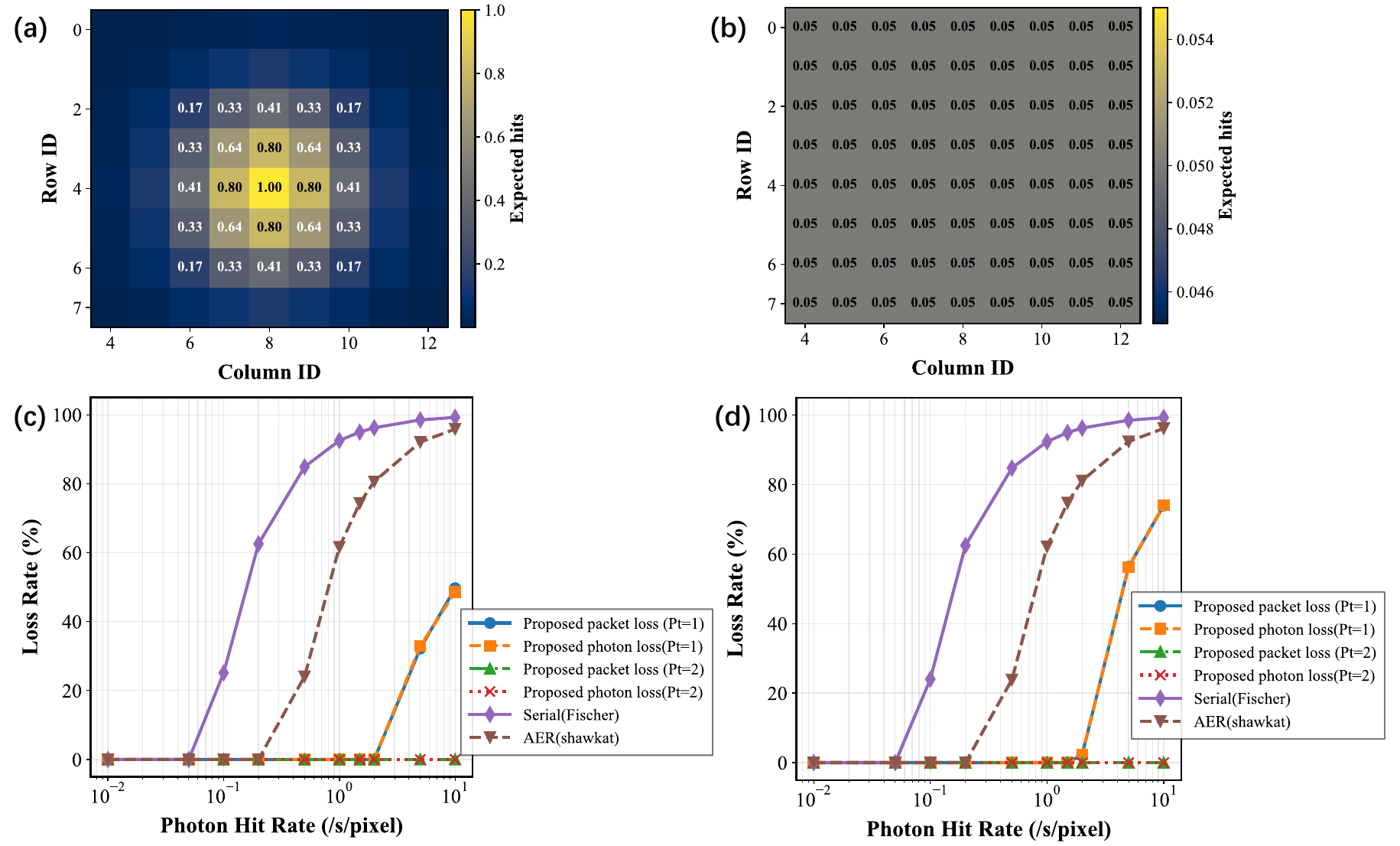}
  \caption{Quantitative evaluation of system-level data loss under distinct spatial photon distributions. Illumination profiles following (a) localized Gaussian and (b) spatially uniform, with (c) and (d) showing their packet and photon loss rates of DarkFlow compared against other architectures.}
  \label{fig:lossrate}
  \vspace{-6pt}
\end{figure}

To evaluate the scalability of the proposed readout system under high photon flux, we develop a custom Python-based simulator. This simulator models the behavioral characteristics and bus arbitration mechanisms of DarkFlow, alongside two existing readout schemes (a serial stream architecture~\cite{fischer2025design} and an AER scheme~\cite{9109635}). In the simulator, photon arrivals are modeled as independent Poisson processes. We generate both localized Gaussian illumination profiles and spatially uniform illumination, shown in Figure~\ref{fig:lossrate}(a) and (b) respectively, with varying total photon rates to emulate different detection scenarios. While macro-scale physics events often exhibit Gaussian distributions, the photon flux incident on a localized sub-array tile is closely approximated by a uniform distribution, which represents the most severe stress test for the readout bus. To ensure a fair comparison, all evaluated architectures are normalized to a uniform $16 \times 8$ SPAD tile array and constrained by a global readout bandwidth of 2.0 Gbps, which represents a typical bottleneck for multi-channel serial I/O interfaces (e.g., LVDS) in modern SiPM implementations~\cite{chen2017mutrig}.

\subsection{Evaluation of System Readout Congestion}

\begin{table*}[t!]
  \centering
  % \small
  \footnotesize
  \setlength{\tabcolsep}{3pt}
  \renewcommand{\arraystretch}{0.85}
  \caption{Comparison of SiPM Readout Architectures}
  \vspace{-6pt}
  \label{tab:architecture_comparison}
  \begin{tabular}{p{3.5cm} p{2.9cm} p{2.9cm} p{2.9cm} p{2.9cm}}
    \toprule
    \textbf{Metric} & \textbf{Gallice~\cite{9459995}} & \textbf{Fischer~\cite{fischer2025design}} & \textbf{Shawkat~\cite{9109635}} & \textbf{Proposed work} \\
    \addlinespace
     & Analog & Serial & AER & DarkFlow \\
    \midrule
    \textbf{Timing representation} & Analog pulse shape & Per-trigger digital stream & Per-trigger address event & Relative-time packet with global anchor \\
    \addlinespace
    \textbf{Native sensing granularity} & $\sim$192{,}000-SPAD macrocell & 9-SPAD tile & 36-SPAD tile & 256-SPAD tile (L2) \\
    \addlinespace
    \textbf{Behavior at high flux} & Dynamic-range limited & Severe bus saturation & Arbitration saturation & Low transport loss in high flux range \\
    \addlinespace
    \textbf{Transport organization} & Analog output & Serialized digital stream & Asynchronous arbitration & Synchronous hierarchical dataflow \\
    \bottomrule
  \end{tabular}
\end{table*}

For DarkFlow, we analyze system performance across different L1 programmable photon thresholds $P_t$. As detailed previously, $P_t$ acts as a hardware-level zero-suppression gate (e.g., $P_t=1$ for single-photon sensitivity vs. $P_t=2$ for dark-count suppression), which directly regulates the baseline event rate and prevents isolated noise from saturating the downstream FIFOs.
Accordingly, the congestion study evaluates packet preservation and photon-equivalent information retention after this front-end encoding.
To evaluate how this compression affects system performance under congestion, we quantify two distinct metrics, standard packet loss that reflects digital bus congestion and photon-equivalent loss that reflects the actual physical energy discarded. Because the L1 stage compresses localized SPAD avalanches into a compact 2-bit energy code, the simulator assigns a physical photon-equivalent weight $w_i$ to each state. By calculating the sum of discarded weights rather than merely counting dropped packets, this provides a fair comparison of physical information preserved across all architectures.

Derived from the aforementioned custom Python-based simulator, Figure~\ref{fig:lossrate}(c) and Figure~\ref{fig:lossrate}(d) compare the congestion robustness and information preservation of each architecture. These results evaluate the packet loss and photon-equivalent loss as functions of the aggregate photon hit rate. Notably, a typical S2 avalanche in dual-phase TPCs generates a peak macroscopic flux ranging from $\mathbf{10^9}$ to $\mathbf{10^{10}}$ hits/s~\cite{agnes2023sensitivity}. We observe that the serial and AER architectures experience severe packet loss, exceeding 60\% at this critical $\mathbf{10^9}$ hits/s threshold and surpassing 80\% at $\mathbf{2 \times 10^9}$ hits/s. Because these architectures generate an independent bus transaction for every SPAD tile trigger, their global arbitration logic saturates quickly at such massive event rates. In contrast, DarkFlow's L1 analog summation and L2 aggregation encode high-frequency avalanches into deterministic packets. With $P_t=1$, DarkFlow maintains ultra-low information loss up to an event rate of $\mathbf{2 \times 10^9}$ hits/s. With $P_t=2$, this lossless operating range extends beyond $\mathbf{10^{10}}$ hits/s.

Table~\ref{tab:architecture_comparison} compares the native sensing granularity, timing representation, transport organization, and high-flux behavior of representative SiPM readout architectures. Serial and AER readouts retain finer native trigger granularity, but under the evaluated bandwidth constraint their transport loss rises sharply at high flux, which limits the spatial and timing information that remains observable at the output. DarkFlow uses coarser frontend grouping, but its hierarchical synchronous transport path preserves spatial tagging and relative timing with low transport loss in the evaluated regime. Compared with analog readout architectures~\cite{9459995}, DarkFlow also provides a fully digital output path that avoids analog pulse clipping under high-flux conditions.

\subsection{Efficiency of Occupancy-Aware Refresh}

\begin{figure}[b!]
  \centering
  \includegraphics[width=\linewidth]{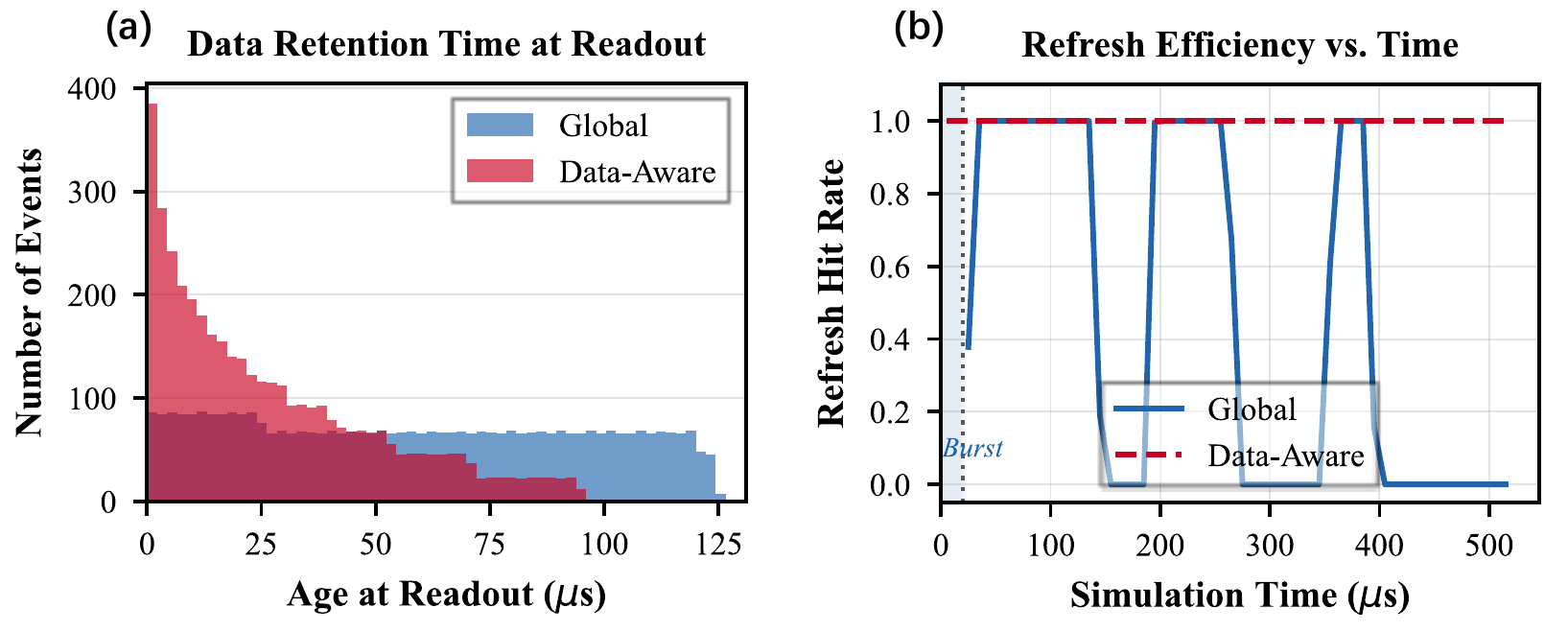}
  \caption{eDRAM refresh efficiency under burst conditions. (a) Comparison of data retention distributions before readout and (b) real-time refresh hit rate.}
  % \vspace{-8pt}
  \label{fig:refresh_performance}
  % \vspace{-4ex}
  \vspace{-6pt}
\end{figure}

To evaluate the proposed occupancy-aware refresh against a conventional global refresh, we simulate the L3 eDRAM FIFO under a worst-case S2 burst condition of 2 billion photons per second per SPAD, using a custom Python-based behavioral model that fully captures the occupancy-aware refresh mechanism. A conventional global refresh sequentially scans all memory rows at fixed intervals, including empty ones. The analysis covers the 0 to 512 $\mathrm{\mu s}$ window from the start of the S2 burst until the FIFO is fully drained via the 2.0-Gbps external link.
We first examine how each refresh strategy affects data age at the time of readout, as shown in Figure~\ref{fig:refresh_performance}(a). Because the occupancy-aware refresh pointer stays within the occupied address range bounded by the read and write pointers, every refresh cycle targets valid data. This produces an age distribution concentrated near zero, meaning most data is refreshed shortly before being read out. We define refresh efficiency as the ratio of valid-data refreshes to total refresh cycles. 
%For this study, we assume a nominal eDRAM retention time of 250 $\mu$s~\cite{8356248}, so data must be refreshed within this window to prevent corruption.
For this study, the refresh analysis uses a nominal eDRAM retention time of 250 $\mu$s as a modeling assumption~\cite{8356248}. Under this assumption, data must be refreshed within this window to avoid corruption.
As shown in Figure~\ref{fig:refresh_performance}(b), the global refresh yields an average hit rate of only 46.8\%, so over half of its cycles are wasted on empty addresses. In contrast, our occupancy-aware refresh maintains a 100\% hit rate throughout the draining window. 
Because every cycle targets valid data rather than empty addresses, DarkFlow achieves a 2.14$\times$ improvement in effective refresh rate, increasing the refresh margin under the assumed retention window.

\subsection{Hardware Implementation Evaluation}

To evaluate the hardware overhead of DarkFlow, we implement the architecture in the GlobalFoundries 22nm process operating at 500 MHz to maintain 2-ns timestamp resolution. The main components of DarkFlow core include the L1/L2 aggregation logic, skid buffers, L3 packing engines, and the eDRAM FIFO controller. 
As detailed in Table~\ref{tab:ppa_breakdown}, the complete DarkFlow architecture occupies 0.255 mm$^2$ and consumes 78.7 mW.
For comparison, a 256$\times$128 SPAD array with a typical 30 $\mu$m pitch occupies approximately 29.5 mm$^2$~\cite{vinayaka2019segmented,vinayaka2019monolithic}. 
Therefore, the DarkFlow digital logic accounts for less than 0.86\% of the detector area, leaving substantial room for physical integration alongside the sensor array.
The area analysis shows that the eDRAM burst-absorbing FIFO dominates total area at 60.6\%, while the entire digital readout datapath accounts for 39.4\%. Other components such as analog local summation, comparison, threshold control, and clock routing only occupy 0.1\% of total area. For applications with shorter burst durations or higher external bandwidth, the eDRAM can be reduced proportionally to further lower the total area.

\begin{table}[t]
  \centering
  % \footnotesize
  \scriptsize
  \setlength{\tabcolsep}{3pt}
  \renewcommand{\arraystretch}{0.85}
  \caption{Area and Power Breakdown of DarkFlow}
  \vspace{-6pt}
  \label{tab:ppa_breakdown}
  \resizebox{\columnwidth}{!}{
  \begin{tabular}{@{}l rr r rr r@{}}
    \toprule
    & \multicolumn{3}{c}{\textbf{Area ($\mu m^2$)}} & \multicolumn{3}{c}{\textbf{Power (mW)}} \\
    \cmidrule(r){2-4} \cmidrule(l){5-7}
    \textbf{Hier. / Module} & \textbf{Val.} & \textbf{Sub\%} & \textbf{Tot\%} & \textbf{Val.} & \textbf{Sub\%} & \textbf{Tot\%} \\
    \midrule
    \textbf{DarkFlow (Top)} & \textbf{255,010} & --- & \textbf{100.0} & \textbf{78.70} & --- & \textbf{100.0} \\
    \midrule
    \hspace{2mm}\textbf{L3 Subsystem} & 100,557 & 39.4 & 39.4 & 49.80 & 63.3 & 63.3 \\
    \hspace{4mm}\textbf{ROW 0} & 12,292 & 12.2 & 4.8 & 6.03 & 12.1 & 7.7 \\
    \hspace{6mm}\textbf{L2 0} & 732 & 6.0 & 0.3 & 0.367 & 6.1 & 0.5 \\
    \hspace{8mm}FIFO & 444 & 60.6 & 0.2 & 0.173 & 47.1 & 0.2 \\
    \hspace{8mm}Event Counter & 101 & 13.8 & <.1 & 0.063 & 17.2 & 0.1 \\
    \hspace{8mm}Skid Buffer & 41 & 5.7 & <.1 & 0.002 & 0.4 & <.1 \\
    \hspace{8mm}Fine Timer & 28 & 3.9 & <.1 & 0.022 & 6.0 & <.1 \\
    \hspace{8mm}Others & 117 & 16.0 & 0.1 & 0.107 & 29.2 & 0.2 \\
    \hspace{6mm}L2 1--15 & 11,560 & 94.0 & 4.5 & 5.663 & 93.9 & 7.2 \\
    \hspace{4mm}Pack \& Timer & 2,097 & 2.1 & 0.8 & 1.18 & 2.4 & 1.5 \\
    \hspace{4mm}ROW 1--7 & 86,108 & 85.6 & 33.8 & 42.54 & 85.4 & 54.1 \\
    % \addlinespace
    \hspace{2mm}\textbf{L3 eDRAM FIFO} & 154,452.5 & 60.6 & 60.6 & 28.90 & 36.7 & 36.7 \\
    \hspace{4mm}eDRAM bank & 153,600.0 & 99.4 & 60.2 & 28.50 & 98.6 & 36.2 \\
    \hspace{4mm}FIFO Logic Ctrl & 852.5 & 0.6 & 0.3 & 0.40 & 1.4 & 0.5 \\
    % \hspace{4mm}Arbiter FSM & 17 & <.1 & <.1 & 4.11 & 14.2 & 5.2 \\
    % \hspace{4mm}Ptr \& Tmr & 130 & 0.1 & <.1 & 0.035 & 0.1 & <.1 \\
    \bottomrule
    \addlinespace
    %\multicolumn{7}{@{}l@{}}{ * \textit{Others}: std cell overheads \& clock routing in L2 0.}
  \end{tabular}%
  }
  % \vspace{-4ex}
  \vspace{-4pt}
\end{table}

Within the digital datapath, several results validate the efficiency of the proposed architecture. Each L2 node costs only 732 $\mu$m$^2$ and 0.367 mW, establishing a clear per-channel cost for scaling to different array sizes. The skid buffers that prevent data loss during stall events consume just 5.7\% of L2 area and 0.4\% of L2 power. This confirms that the distributed buffering in the systolic readout adds negligible overhead. Similarly, the FIFO logic controller that implements the occupancy-aware refresh occupies only 0.6\% of total DRAM area and 1.4\% of DRAM power, so the 2.14$\times$ refresh efficiency improvement comes at minimal hardware cost. 

On the power side, the digital logic consumes 63.3\% of total power while the DRAM accounts for 36.7\%. When distributed across the sensor substrate area of approximately 30 mm$^2$, the total dissipation yields a thermal density of roughly 2.6 mW/mm$^2$, which complies with the thermal budget reported for immersed detector electronics in liquid argon environments~\cite{consiglio2020cryogenic}.

\vspace{-2pt}
\section{Conclusion}
This work presents DarkFlow, a hierarchical digital SiPM readout architecture that unifies sparse S1 and high-flux S2 signal modalities within a single deterministic dataflow framework. The architecture features local aggregation, compact hierarchical packet formatting, distributed backpressure, and eDRAM burst buffering with occupancy-aware refresh to address the conflicting demands of nanosecond-level timing, low data loss, and high throughput in large-scale dark matter detectors. 
We demonstrate that DarkFlow maintains ultra-low packet loss at billion-photon event rates, where conventional serial and AER architectures exceed 80\% data loss. The occupancy-aware refresh achieves a 2.14$\times$ improvement in effective refresh rate by targeting only valid addresses. The entire DarkFlow architecture occupies 0.255 mm$^2$ and consumes 78.7 mW in GF 22nm, accounting for 0.86\% of the total detector area and complying with the power budget in the liquid argon environment.

\vspace{-4pt}
\begin{acks}
\footnotesize
This work was partially supported by the Department of Energy (DOE) Office of Science, Office of High Energy Physics grant number DE-SC0022313 awarded to the University of California to support the HEPCAT consortium and by DE-SC0025540.

\end{acks}

\vspace{-4pt}
\bibliographystyle{ACM-Reference-Format}
\bibliography{glsvlsi-2026}

% \appendix









\end{document}